\documentclass[12pt]{article}
\usepackage{amsmath}
\usepackage{latexsym}
\usepackage{bbm}

\def\a{\alpha}

\def\b{\beta}
\def\g{\gamma}

\def\d{\delta}

\def\e{\varepsilon}

\def\th{\theta}

\def\l{\lambda}
\def\m{\mu}
\def\n{\nu}

\def\r{\rho}

\def\s{\sigma}
\def\t{\tau}

\def\D{\Delta}

\def\O{\Omega}

\def\pa{\partial}

\def\half{\frac{1}{2}}

\def\tr{{\rm tr}}

\def\and{{\rm and}}

\def\ie{{\it i.e.,} }

\def\IR{\mathbbm R}
\def\IZ{\mathbbm Z}

\begin{document}
\vspace*{-1.0in}
\thispagestyle{empty}
\begin{flushright}
CALT-68-2867
\end{flushright}
\baselineskip = 18pt
\parskip = 6pt

\vspace{1.0in}

{\Large
\begin{center}
Highly Effective Actions
\end{center}}

\vspace{.25in}

\begin{center}
John H. Schwarz\footnote{jhs@theory.caltech.edu}
\\
\emph{California Institute of Technology\\ Pasadena, CA  91125, USA}
\end{center}
\vspace{.25in}

\begin{center}
\textbf{Abstract}
\end{center}
\begin{quotation}
\noindent

It is conjectured that the world-volume action of a probe D3-brane in an ${AdS_5 \times S^5}$
background of type IIB superstring theory, with one unit of flux, can be
reinterpreted as the {\em exact} effective action (or highly effective action)
for $U(2)$ ${\cal N} =4$ super Yang-Mills theory on the Coulomb branch. An analogous
conjecture for $U(2)_k \times U(2)_{-k}$ ABJM theory
is also presented. The main evidence supporting these conjectures is that
the brane actions have all of the expected symmetries and dualities. Highly effective
actions have general coordinate invariance, even though they describe nongravitational
theories.

\end{quotation}

\newpage

\pagenumbering{arabic}

\tableofcontents

\newpage

\section{Introduction}

In the opening lecture at the Strings 2012 conference in Munich, I discussed some lessons
learned in the course of my career. The one that is relevant here is: ``Take
coincidences seriously.'' This principle has served me well on at least one
previous occasion \cite{Scherk:1974ca}. However, there have also been missed opportunities.
For example, many of us were well aware of
the fact that the isometry group of AdS space in $d+1$ dimensions is the same as the conformal
group in $d$ dimensions. This was generally assumed to be a strange, but surely irrelevant,
coincidence. We now know better. In fact, the AdS/CFT correspondence will play a central
role in the discussion that follows. However, the point of this paper is to emphasize
a different coincidence and to suggest that it should be taken seriously.

This paper examines the construction of the world-volume theory of a single isolated $p$-brane
in an $AdS_{p+2} \times S^n$ background with $N$ units of flux threading the sphere.
The specific examples studied are the ones with maximal supersymmetry as well as
ABJM theory, which has 3/4 maximal supersymmetry.
The actions are of the usual type, consisting of a sum of two terms,
$S_1 + S_2$, which is sometimes denoted $S_{DBI} + S_{WZ}$.
The approximations that are implicit in probe-brane constructions are well-known. One of them is
the probe approximation in which the effects of the probe brane on the background geometry
and the gauge field configuration are neglected. This is tantamount to a large-$N$ approximation.
The other approximation is the assumption that world-volume fields,
for example a Born--Infeld $U(1)$ field strength, are slowly varying.
This justifies excluding consideration of possible terms involving
higher derivatives of fields. The field strength itself is allowed to be large.
Despite these approximations, the formulas that are obtained in these
constructions have a beautiful property: they fully incorporate the symmetry of the background
as an exact global symmetry of the world-volume theory.
This symmetry is the superconformal group $PSU(2,2|4)$ in the case of a
D3-brane in $AdS_{5} \times S^5$, for example. In this example it also includes
the $SL(2,\IZ)$ duality group, which is known to be an exact symmetry of type IIB
superstring theory.

There was quite a bit of activity studying world-volume actions for branes in
these geometries at the end of the last century. However, most of that work
focused on superstrings, rather than higher-dimensional branes. An important example is the
superstring in $AdS_{5} \times S^5$, which was worked out in \cite{Metsaev:1998it}.
Also, those works that did
study $p$-branes in $AdS_{p+2}$ had different motivations from ours, and therefore
made coordinate and gauge choices that are different from the ones made here. Our choices
are specifically tuned to a particular goal: presenting the world-volume action of the
brane in a form in which it can be interpreted as a candidate solution to an
entirely different problem. That problem is the construction of the {\em effective
action for a superconformal field theory on the Coulomb branch}. The fact that
the world-volume theory of a $p$-brane in an $AdS$ background is conformally invariant
has been frequently noted, for example in \cite{Maldacena:1997re} -- \cite{Bellucci:2002ji}.
The proposal that the result can be reinterpreted in the manner presented here motivates analyzing
these brane actions in a very specific manner.

Let us discuss $U(2)$ ${\cal N} =4$ super Yang--Mills theory to be specific. For most
purposes it is correct to ignore a decoupled abelian multiplet and to speak of $SU(N)$
rather than $U(N)$. However, $U(N)$ is important for obtaining an $SL(2, \IZ)$ duality
group.\footnote{I am grateful to N. Seiberg for a discussion. See \cite{Aharony:2013hda}
for an explanation of the relevant issues.}
On the Coulomb branch the ``photon'' supermultiplet remains massless and the ``$W^{\pm}$''
supermultiplets acquire a mass. In principle, the effective action on the Coulomb branch
is obtained by performing the path integral over the massive fields,
thereby ``integrating them out'' and
producing a very complicated formula in terms of the massless photon supermultiplet only.
If the computation could be done exactly, the resulting effective action would still
encode the entire theory on the Coulomb branch.
It would {\em not} be just a low-energy effective action. We propose
to call such an effective action a {\em highly effective action} (HEA).
Clearly, we do not have the tools to carry out such an exact computation,
but in some cases (such as the one mentioned above) we do know many of the properties
that the HEA should possess.

An important fact about the HEA is that it has all of the global symmetries of the
original theory, though some of them are spontaneously broken. This is known to be
true in the formulation with explicit $W$ supermultiplets. It should continue to be true
after they are integrated out, especially if one has the complete answer. The symmetry breaking
is simply a consequence of assigning a vacuum expectation value (vev) to a massless scalar
field.\footnote{This field has sometimes been called
the ``dilaton.'' We prefer not to use that terminology, because
it is unrelated to the string theory dilaton, which controls the coupling constant.
In fact, in a brane construction this scalar field corresponds to the radial
coordinate of the AdS space.}
As this vev goes to zero, the multiplets that have been integrated out become massless
and the formulas become singular. However, one does not need to specify the vev, so the
action realizes the full symmetry. There are really only two discrete options, since the vev is
the only scale in the problem. Either it is zero, or it can be set to one. In principle, the
HEA contains all information about the theory, at energies both below
and above the scale set by the vev. What one can
question, indeed should question, is whether the procedure described in this paper gives
the correct formula.

In the (known) formulation of the Coulomb branch theory, in which the massive
$W^{\pm}$ fields have not been integrated out, there are other massive particles (monopoles
and dyons) that do not appear explicitly in the action. Yet the theory is fully specified
at all scales. After the $W^{\pm}$ have been integrated out, they should be on an equal footing
with the monopoles and dyons, and one still has an exact characterization of the theory.
In fact, the HEA should have soliton solutions, analogous to
those considered in \cite{Callan:1997kz} (dubbed {\em BIons} in \cite{Gibbons:1997xz}),
that form complete $SL(2, \IZ)$ multiplets. One of the advantages of the HEA description
of the Coulomb branch is that dualities, such as $SL(2,\IZ)$, are much
easier to understand than in the formulation with explicit $W$ fields. We will argue
that there are other advantages, as well.

Since ${\cal N} =4$ super Yang--Mills theory in the unbroken phase is conformal,
we know that every term in the expansion of the Coulomb-branch effective action
should have dimension four,
which is the spacetime dimension of the field theory. There are no dimensionful parameters
other than the vev, which does not need to be specified.
The only surprising feature of the effective action
(if you haven't thought about this before)
is that inverse powers of a scalar field appear.
Thus, individual terms can be arbitrarily complicated and yet have their
dimensions end up as four simply by including appropriate (inverse) powers
of the scalar field.

The action that we will obtain for a suitably placed $p$-brane in a
background geometry containing an $AdS_{p+2}$ factor has
all the symmetries and other properties that a Coulomb branch effective action
should have. Therefore, this action provides a compelling candidate
for the HEA. However, there are some caveats. First, it is not obvious
to what extent symmetries and other properties determine the solution. So this could
be a wrong formula that just happens to have many correct properties, thereby
demonstrating that taking coincidences too seriously can lead you astray.
The general expectation is that the larger the superconformal symmetry group that the
action should incorporate, the more limited the possibilities are.
When there are additional requirements, such as dualities,\footnote{We do not call
the $SL(2, \IZ)$ dualities `symmetries,' because they relate the theory at different values of
the coupling constants. In some settings, they even relate different theories. $SL(2, \IZ)$ is
a symmetry of type IIB superstring theory, which is spontaneously broken by the choice of vacuum.
For specific values of the modulus a $\IZ_2$ or $\IZ_3$ subgroup can survive as a symmetry.}
they can greatly strengthen the case. A second caveat is that the solutions that
are obtained by brane constructions depend on an
integer parameter, $N$, which is the number of units of background flux through
the sphere. We will find that the brane action satisfies all of the symmetry
and duality requirements for any
choice of $N$, so there is at least this much nonuniqueness. Even though
the probe approximation is only valid for large $N$, the choice $N=1$ is the most natural
candidate for the HEA of the $U(2)$ gauge theory. We will ignore the
decoupled $U(1)$ multiplet, which was discussed earlier. It needs to be
adjoined to the brane-probe action.

This paper only discusses the bosonic degrees of freedom.
This allows us to emphasize conceptual issues with a
minimum of distracting technicalities. Also, it is important to have these formulas
completely debugged, before confronting the more complicated formulas with
fermions. The ultimate inclusion of the fermi fields is
essential for the results to be truly meaningful, so that is an important project for
the future. However, the formulas presented here already are of interest,
since they are the bosonic truncations of the HEAs that include fermi
fields. Some of the formulas presented here, such as the bosonic part of the D3-brane
action, have appeared previously in the literature. However, we have clarified a few details,
and set the stage for inclusion of the fermions in a form appropriate for reinterpretation
as a Coulomb-branch effective action.


The local symmetries of the world-volume theory, which are general coordinate invariance
and kappa symmetry (when fermions are included), provide crucial constraints in the construction of
$p$-brane world-volume actions. There is a natural gauge choice that brings the gauge-invariant formulas into a recognizable
form containing only the expected supermultiplet of fields.
The local symmetries are no longer apparent in the gauge-fixed theory, but
they control some of the symmetry transformations of the resulting theory.
The supersymmetry transformations, for example, take a rather simple form before
gauge fixing. They become much more complicated after gauge fixing, because they then
include compensating local kappa transformations required to maintain the gauge choice.
More mundane examples of this are already present in the bosonic truncations described
in this paper. Thus, an understanding of the local symmetries gives a systematic procedure
to derive how symmetry transformations of the HEA incorporate quantum effects
attributable to the fields that have been integrated out.
The fact that these nongravitational theories are naturally
formulated with general coordinate invariance is quite intriguing.

Section 2 describes the construction of the HEA that corresponds to the
world-volume theory of a probe D3-brane in $AdS_5 \times S^5$. Section 3
describes analogous M2-brane and D2-brane constructions of the HEA for
$U(2)_k \times U(2)_{-k}$ ABJM theory on the Coulomb branch. The two descriptions
are shown to be related by a duality transformation, analogous to the S-duality
of the four-dimensional theory of Section 2, which proves their equivalence.
Section 4 describes the M5-brane in $AdS_7 \times S^4$. The analysis
in each case utilizes the Poincar\'e-patch description of anti de Sitter
spacetime. The Poincar\'e patch is reviewed in Appendix A.
Appendix B demonstrates the S-duality of the D3-brane HEA.
Appendix C discusses an alternative to the Poincar\'e patch,
namely the geodesically complete covering space,
and explains why it is not appropriate for our purposes.

\section{The D3-brane in $\mathbf {AdS_5 \times S^5}$}

Type IIB superstring theory has a maximally supersymmetric solution with $PSU(2,2|4)$
isometry. In this paper we only consider bosonic of freedom, so the relevant part of the symmetry
is the bosonic subgroup $SO(4,2) \times SO(6)$. The solution has the ten-dimensional geometry
${AdS_5 \times S^5}$, where both factors have radius $R$. Also, $N$ units of five-form flux
$F_5= dC_4$ thread the five-sphere. Since  $F_5$ is self dual, it
follows that it is proportional to ${\rm vol} (AdS_5) + {\rm vol} (S^5)$, with a constant of
proportionality to be discussed later.

Using the Poincar\'e patch coordinates discussed in Appendix A,
the ten-dimensional metric is
\begin{equation}
ds^2 = R^2 \left( v^2 dx \cdot dx + v^{-2} dv^2 + d\O_5^2 \right)
= R^2\left(v^2 dx\cdot dx + v^{-2} dv \cdot dv\right).
\end{equation}
$v$ is now the length of the six-vector $v^I$, \ie $v^2 = \sum (v^I)^2 = v \cdot v$.
This is an $SO(6)$ invariant inner product. We can also check the conformal
symmetry of the $AdS_5$ volume form
\begin{equation}
{\rm vol} (AdS_5) = R^5 e^0\wedge e^1 \wedge e^2 \wedge e^3 \wedge f .
\end{equation}
Using Eqs.~(\ref{deltae}) and (\ref{deltaf}), it is clear that $\d f$ gives
a vanishing contribution to $\d {\rm vol} (AdS_5)$ and that the contribution
of $\d e^\m$ is proportional to $\tr (b^\m x_\n - x^\m b_\n) =0$.

The coordinates $v^I$ will be identified as the
world-volume scalar fields of the D3-brane.
However, as introduced here, they would not be nicely
normalized. Therefore, let us define $v^I = \sqrt{c_3} \, \phi^I$,
where $c_3$ is a dimensionless constant
that will be chosen later to give the desired normalization. The metric becomes
\begin{equation}
ds^2 = g_{MN}dx^M dx^N = R^2\left(c_3 \, \phi^2 dx\cdot dx + \phi^{-2} d\phi \cdot d\phi \right).
\end{equation}
where $x^M = (x^\m, \phi^I)$, $g_{\m\n} = c_3 R^2 \phi^2\eta_{\m\n}$, $g_{IJ} = R^2 \phi^{-2} \d_{IJ}$.
In these coordinates $\phi = \infty$ is the boundary of AdS and $\phi = 0$ is the Poincar\'e-patch
horizon.

The radius $R$, raised to the fourth power, is determined by the ten-dimensional type IIB
superstring theory solution to be
\begin{equation} \label{R4}
R^4 = 4\pi g_s N (\a')^2 .
\end{equation}
The Regge slope $\a'$ is given by $\a' =l_s^2$, where $l_s$ is the string length scale.
Here we are interested in ten dimensions. However,
in $D$ dimensions Newton's constant, $G_D$, is given by
$32 \pi^2 G_D = (2\pi l_p)^{D-2} = g_s^2 (2\pi l_s)^{D-2}$,
where $l_p$ is the $D$-dimensional Planck length.

The D3-brane world-volume action is given as the sum of two terms
$S = S_1 + S_2$. $S_1$ has a Dirac/Born--Infeld/Nambu--Goto type structure. When fermions are
included (for the case of a flat-spacetime background), it also has a Volkov--Akulov
structure. $S_2$ has a Chern--Simons/Wess--Zumino
type structure. \footnote{Moreover, when fermions are included, $S_1$ and $S_2$
are related by local kappa symmetry, which was discovered for the massless superparticle
of \cite{Brink:1981nb}, which has no $S_2$ term, in \cite{Siegel:1983hh} and utilized for the superstring, which does have an $S_2$ term, in \cite{Green:1983wt}.}

The standard formula for the bosonic part of $S_1$ for a D-brane is a functional
of the embedding functions
$x^M(\s^\a)$ and a world-volume $U(1)$ gauge field $A_\b (\s^\a)$ with field strength
$F_{\a\b} = \pa_\a A_\b - \pa_\b A_\a$:
\begin{equation} \label{S1}
S_1 = -T_{D3} \int \sqrt{- \det \left(G_{\a\b} + 2\pi \a' F_{\a\b}\right)}\, d^4 \s ,
\end{equation}
where the D3-brane tension is
\begin{equation} \label{D3tension}
T_{D3} = \frac{2 \pi}{g_s(2\pi l_s)^4} .
\end{equation}
Combining Eqs.~(\ref{R4}) and (\ref{D3tension}),
\begin{equation} \label{R4T}
R^4 T_{D3} = \frac{N}{2\pi^2} .
\end{equation}
Also,
\begin{equation}
2\pi \a'/R^2 = \sqrt{\pi/ g_s N}.
\end{equation}

In general, the $S_1$ integrand would contain a factor $e^{-\Phi}$, where $\Phi$ is
the ten-dimensional dilaton field. However, in the background under consideration this factor
is a constant, $1/g_s$, which is included in $T_{D3}$. $G_{\a\b}$ is the induced
four-dimensional world-volume metric
\begin{equation}
G_{\a\b} = g_{MN}(x) \pa_\a x^M \pa_\b x^N.
\end{equation}
The action in Eq.~(\ref{S1}) has four-dimensional general coordinate invariance, since the
integrand transforms as a scalar density. Furthermore, the conformal symmetry group is realized
as a global symmetry, since $G_{\a\b}$ and $F_{\a\b}$ are separately invariant under the entire
conformal group. The previous analysis ensures this for $G_{\a\b}$. In the case of $F_{\a\b}$
it is a triviality: the gauge field $A_\a$ is inert under the entire conformal group. It will
transform nontrivially after a gauge choice is made.

Substituting the ten-dimensional metric $g_{MN}$ given previously, pulling out some factors
from the square root, and using Eqs.~(\ref{R4}) and (\ref{R4T}), we obtain
\begin{equation} \label{S1a}
S_1 = -\frac{N}{2\pi^2} \int  \phi^4 \sqrt{- \det \left(c_3 \, \pa_\a x \cdot\pa_\b x
+ \phi^{-4} \pa_\a \phi \cdot \pa_\b \phi + \sqrt{{\pi}/(g_s N)}  \phi^{-2} F_{\a\b}\right)}\, d^4 \s.
\end{equation}
The next step is to use the general coordinate invariance symmetry to
choose a convenient gauge. The natural
and appropriate choice for our purposes is static gauge. This gauge
identifies the world-volume coordinates $\s^\a$ with the spacetime coordinates $x^\m$:
\begin{equation}
x^\m(\s) = \d^\m_\a \s^\a.
\end{equation}
Then the action takes the form
\begin{equation} \label{S1new}
S_1 = -\frac{Nc_3^2}{2\pi^2} \int  \phi^4 \sqrt{- \det \left( \eta_{\m\n}
+ \frac{\pa_\m \phi \cdot \pa_\n \phi}{c_3\, \phi^4} +  \sqrt{{\pi}/(g_s N)}
\frac{F_{\m\n}}{c_3\, \phi^2}\right)} \, d^4 x.
\end{equation}
Having chosen the static gauge, the fields $\phi^I$ and $A_\m$ become functions of
$x^\m$. The four-dimensional spacetime metric is simply the flat Minkowski metric denoted
$\eta_{\m\n}$.

If one expands out the square root in Eq.~(\ref{S1new}) in powers of $\pa \phi$ and $F$,
the leading term is proportional to $\int \phi^4 \, d^4 x $. In his famous
paper \cite{Maldacena:1997re}, Maldacena explained that this term must
be canceled, since such a term would imply that a force acts on the brane in the radial
direction. We know that we are dealing with a system of parallel BPS branes at rest,
for which there must be a perfect cancellation of forces. Thus, even though this term is
conformally invariant and $SO(6)$ invariant, it should not appear. We will demonstrate below
that it cancels against a contribution from $S_2$. So the requisite cancellation will
arise naturally without any need to introduce it in an {\it ad hoc} manner.

The next terms in the expansion of the Lagrangian are
the kinetic terms of the free theory
\begin{equation}
L_{\rm free} = -\frac{Nc_3}{4\pi^2}\pa_\m \phi \cdot \pa^\m \phi - \frac{1}{8\pi g_s}F_{\m\n} F^{\m\n}.
\end{equation}
Requiring that $\phi^I$ is canonically normalized leads to the choice
$c_3 = 2 \pi^2/N$. A better alternative is
\begin{equation}
c_3 = \frac{\pi}{g_s N}.
\end{equation}
For this choice the entire action depends only on the 't Hooft parameter
$\l = g_s N$, aside from an overall factor of $N$ (or $1/g_s$). This structure
suggests the conjecture that the loop expansion of this action (after we include $S_2$ and fermions)
corresponds to the string theory loop expansion, or equivalently to
the genus (or $1/N$) expansion at fixed $\l$.  This conjecture is somewhat uncertain, because
we have not proposed a precise interpretation of the formula with $N>1$.
We will comment on this issue in the conclusion.

The conjecture that the loop expansion of the $N=1$ HEA corresponds to the topological
expansion of the super Yang-Mills theory could be wrong even if our main conjecture, namely
that we have found the HEA for the $U(2)$ theory, is correct. However, if this
secondary conjecture is also correct, there would be
interesting consequences. For one thing,
it would imply that the HEA action (after adding the contributions of
fermions and of $S_2$) has dual superconformal symmetry when treated classically, and
hence the full Yangian symmetry.  It should be very interesting to compute scattering
amplitudes in the tree approximation to determine whether they are compatible with this
symmetry. This is a very clean problem, since there are no integrals to evaluate and
no infrared divergences to regulate. There could be beautiful formulas waiting to be discovered.
One could also explore Wilson loops.

An important fact is that the construction described here
ensures that the action is conformally invariant and $SO(6)$ invariant.
(When fermions are included the full $PSU(2,2|4)$ superconformal symmetry will be built in.)
The conformal symmetry and the $SO(6)$ symmetry are both
spontaneously broken when one assigns a nonzero expectation value $\langle \phi^I \rangle$.
This ensures that the inverse powers of $\phi$ are well-defined. Physically,
$\langle \phi^I \rangle$ describes the position of the brane in the radial
direction of $AdA_5$ and on the $S^5$. In the limit that $\langle \phi \rangle$ goes to zero,
the brane approaches the horizon, and new massless
degrees of freedom arise. This is the significance of the singularities at $\phi =0$.

Having fixed the static gauge, the formulas for infinitesimal conformal
transformations are modified by the addition of a compensating general coordinate
transformation, which is required to maintain the gauge choice. Denoting the new
transformation by $\D$ and the old one by $\d$, we have
\begin{equation}\label{Dx}
\D x^\m = \d x^\m + \xi^\m = 0.
\end{equation}
\begin{equation}
\D \phi^I = \d \phi^I +\xi^\m \pa_\m \phi^I
\end{equation}
\begin{equation}
\D A_\m =  \xi^\n F_{\n\m}.
\end{equation}
Equations (\ref{conformal1}) and (\ref{Dx}) give
\begin{equation}
\xi^\m = -  b^\m (\frac{1}{c_3 \, \phi^2} + x\cdot x) + 2 b \cdot x x^\m.
\end{equation}
Hence, with this value of $\xi^\m$, the compensated conformal transformations of the
remaining bosonic fields in the static gauge are
\begin{equation}
\D \phi^I = 2b \cdot x \phi^I + \xi^\m \pa_\m \phi^I
\end{equation}
\begin{equation}
\D A_\m = \xi^\n F_{\n\m}.
\end{equation}
When fermions are included, there will be an analogous analysis of supersymmetry
transformations involving compensating local kappa transformations. From the point of
view of the effective action, these compensating gauge transformation contributions
to global symmetry transformations correspond to quantum corrections that arise from
integrating out the massive $W$ supermultiplets.

The second term in the D3-brane action, $S_2$, is
an integral of a 4-form whose bosonic part has terms of the form
$C_4$, $C_0 F \wedge F$, $C_0 \tr(R \wedge R)$ with coefficients to be discussed later.
The two-forms $B_2$ and $C_2$ do not appear, because they vanish
in the $AdS_5 \times S^5$ background. Here $F$ is the abelian world-volume field strength, which
already appeared in $S_1$. $C_4$ is the RR 4-form whose field strength, $F_5 = d C_4$,
is self dual. $R$ is the ten-dimensional spacetime curvature two-form, pulled
back to the D3-brane world volume. Fortunately, $\tr(R \wedge R)$ vanishes in the
$AdS_5 \times S^5$ background, so we will not discuss it further. $C_0$ is
the RR 0-form, which can take an arbitrary constant value $\langle C_0 \rangle = \chi = \th/2\pi$
in the $AdS_5 \times S^5$ background. The parameter $\th$ will correspond to the usual theta
angle of the gauge theory. The theory is expected to be invariant under $\chi \to \chi +1$,
which is the T transformation of the $SL(2,\IZ)$ duality group.

The self-dual five-form field $F_5$ in the $AdS_5 \times S^5$ background is proportional to the
sum of the volume form for the $S^5$ and its dual, which is the volume form for the $AdS_5$. Thus,
the field strength is
$F_5 = k_3( {\rm vol}(S^5) + {\rm vol} (AdS_5)),$ where the coefficient $k_3$ depends on normalization
conventions. Referring to Eq.~{(\ref{AdSmetric}), we read off the volume
form for the unit radius $AdS_5$ space as the wedge product of five one-forms
\begin{equation}
{\rm vol} (AdS_5) = v^3 dv \wedge dx^0 \wedge dx^1 \wedge dx^2 \wedge dx^3
= d \left( \frac{v^4}{4} dx^0 \wedge \dots \wedge dx^3\right) .
\end{equation}

We can now write the $C_4$ part of $S_2$ in the form
\begin{equation} \label{intC4}
\mu_3 \int_{\pa M} C_4 =  \mu_3 k_3 \int_M ( {\rm vol}(S^5) + {\rm vol} (AdS_5)) =
\mu_3 k_3 \int_M  {\rm vol} (AdS_5).
\end{equation}
Here $M$ denotes a five-dimensional region whose boundary is the D3-brane world volume.
There is a natural choice for this region. Recall that the fields $v^I$ determine the
radial coordinate $v$ and the position on the five-sphere of the D3-brane. If we choose
the region $M$ to be the region between $0$ and $v$, we can pass from the five-form to
the four-form by integrating $v'$ from $0$ to $v$. The upper limit gives the desired result
and the lower limit contributes zero. Thus,
\begin{equation}  \label{S2AdS}
 \mu_3 k_3 \int_M  {\rm vol} (AdS_5) = \frac{\mu_3 k_3}{4} \int_{\pa M} v^4 dx^0 \wedge \dots dx^3
= \frac{\mu_3 k_3 c_3^2}{4} \int \phi^4 d^4 x.
\end{equation}
While this is a correct formula, one should understand that
\begin{equation}
d^4x = \sqrt{-\det(\eta_{\m\n}\pa_\a x^\m \pa_\b x^\n)} d^4 \s.
\end{equation}
In static gauge, the Jacobian factor is equal to one, and $d^4\s = d^4 x$.

The next step is to evaluate $\mu_3 k_3$. The way to do this is to use the fact that there are
$N$ units of five-form flux. This means that the {\em same} five-form, $\mu_3 F_5$,
integrated over an $S^5$ of unit radius is $2\pi N$:
\begin{equation}
\mu_3 k_3 \int_{S^5} ( {\rm vol}(S^5) + {\rm vol} (AdS_5))
= \mu_3 k_3 \int_{S^5} {\rm vol}(S^5) = \mu_3 k_3 \pi^3 = 2 \pi N.
\end{equation}
Substituting $\m_3 k_3 c_3 =4$ in Eq.~(\ref{S2AdS}), and restoring the $\chi$ term, we obtain
\begin{equation}
S_2 =  \frac{Nc_3^2}{2\pi^2} \int \phi^4 d^4 x + \frac{\chi}{8\pi} \int F \wedge F .
\end{equation}
Combining this with $S_1$ in Eq.~(\ref{S1new}) gives the total result.
\begin{equation} \label{finalD3}
S = -\frac{Nc_3^2}{2\pi^2} \int  \phi^4 \left(\sqrt{- \det \left( \eta_{\m\n}
+ \frac{\pa_\m \phi \cdot \pa_\n \phi}{c_3\, \phi^4} +  \sqrt{{\pi\t_2}/ N}
\frac{F_{\m\n}}{c_3\, \phi^2}\right)} -1\right)\, d^4 x
+ \frac{\t_1}{8\pi} \int F \wedge F .
\end{equation}
Here we have introduced the parameter
\begin{equation} \label{taudef}
\t = \t_1 +i\t_2 = \langle C_0 + i \exp(-\Phi) \rangle = \chi + i/g_s.
\end{equation}
For the specific choice $c_3 = \pi/\l$, where $\l = g_s N$,
\begin{equation} \label{finalD3}
S = -\frac{1}{2g_s \l} \int  \phi^4 \left(\sqrt{- \det \left( \eta_{\m\n}
+ \frac{\l}{\pi}\frac{\pa_\m \phi \cdot \pa_\n \phi}{ \phi^4} +  \sqrt{\frac{\l}{\pi}}
\frac{F_{\m\n}}{\phi^2}\right)} -1\right)\, d^4 x
+ \frac{\chi}{8\pi} \int F \wedge F .
\end{equation}

As promised earlier, there is a precise cancellation of the undesired $\int \phi^4 d^4x$ terms,
which arises from combining $S_1$ and $S_2$. Had we considered an anti-D3-brane
(a $\overline{\rm D3}$-brane) instead, $S_2$
would have had the opposite sign, and the cancellation of forces would not have occurred.
Rather, there would be a potential $V \sim \phi^4$, which implies that
the $\overline{\rm D3}$-brane would be attracted to $\phi =0$, which is the Poincar\'e-patch
horizon. In the Coulomb branch interpretation, this is the origin of the moduli
space, the singular point at which fields that have been integrated out become massless,
and the nonabelian gauge symmetry is restored. It is known that a potential term in a
Coulomb-branch effective action is not consistent with ${\cal N} =4$
supersymmetry \cite{Seiberg:1988ur}. It is also known that the world-volume theory of
a $\overline{\rm D3}$-brane probe is not supersymmetric.

In Appendix B we demonstrate that the S-duality group of Eq.~(\ref{finalD3})
is precisely $SL(2, \IZ)$ for all $N$. Under the S transformation of the duality group, the
gauge field $A_\m$ is replaced by a dual gauge field and $\t \to -1/\t$.  No proof is known
for the formulation of the theory with explicit $W$ supermultiplets.
It is truly remarkable that this complicated nonlinear formula, whose essential structure
was introduced by Born and Infeld almost 80 years ago \cite{Born:1934gh},
should have this symmetry. This fact
provides further confirmation that the action derived by studying a probe D3-brane has all
the properties required for the effective action for the $U(2)$ ${\cal N} = 4$
super Yang-Mills theory on the Coulomb branch. The only freedom is the choice of the integer
parameter $N$. The choice $N=1$ is probably the correct one. The actions with $N>1$ may
play a role in the construction of effective field theories for higher-rank gauge groups,
but there must be additional ingredients as well.

As mentioned in the introduction, this HEA should have soliton solutions that form a
complete $SL(2, \IZ)$ multiplet. These will describe the $W$ supermultiplets that have
been integrated out as well as monopole and dyon supermultiplets. From a string theory
viewpoint, these can be interpreted as $(p,q)$ strings ending on the D3-brane. There
should also be instanton solutions of the Euclideanized theory. These can be
interpreted as embedded D(-1)-branes.

Appendix C analyzes an analogous problem in which a D3-brane of $S^3$ topology probes
global ${AdS_5 \times S^5}$. In this case it is shown that the contributions to the potential
from $S_1$ and $S_2$ do not cancel. As a result, the brane probe is unstable to decay
and shrinking to a point at the center of global $AdS_5$. In terms of the
nonabelian gauge theory on $S^3$, the interpretation is that there is no Coulomb branch.
We will not repeat the analysis of this appendix for the brane theories that we
consider in the next sections. Only the Poincar\'e-patch descriptions will
be presented.

\section{ABJM Theory}

The purpose of this section is to derive a candidate formula for the HEA for
$U(2)_k \times U(2)_{-k}$ ABJM theory \cite{Aharony:2008ug} on the Coulomb branch.
A formula will be derived first by considering an M2-brane in 11 dimensions.
Then an alternative formula will be obtained by considering a type IIA D2-brane
in $AdS_4 \times CP^3$. The equivalence of the two formulas will be demonstrated.
The D2-brane version of the
formula has the field content and structure of an abelian ${\cal N} = 6$
super Yang--Mills theory, whereas the M2-brane version
has the field content and structure of a $U(1)_k \times U(1)_{-k}$ ABJM theory \ie an
${\cal N} = 6$ superconformal Chern--Simons theory.
All calculations will be carried out for $N$ units of flux. However, as before, we
conjecture that $N=1$ is the relevant choice.

\subsection{The M2-brane in $\mathbf {AdS_4 \times S^7}$}

This geometry corresponds to the $k=1$ special case of the more general geometry
$AdS_4 \times S^7/\IZ_k$, which is considered in the next subsection.
In the M2-brane case the radius of the $AdS_4$ is half the radius $R$ of the $S^7$,
and the coordinate $x^\m$ is a 3-vector. So, in terms of Poincar\'e-patch coordinates
for the AdS factor, the 11-dimensional metric is
\begin{equation} \label{M2metric}
ds^2 = (R/2)^2 \left( v^2 dx \cdot dx + v^{-2} dv^2\right) + R^2 d\O_7^2
= R^2\left({c_2}\phi^4 dx \cdot dx + \phi^{-2} d\phi^2 + d\O_7^2\right).
\end{equation}
The change of variables $v = 2\sqrt{c_2}\phi^2$ is motivated by
the fact that $v$ has dimension 1, whereas a scalar field in three dimensions
should have dimension 1/2. As in the D3-brane case, the constant $c_2$ is included so that
the normalization of the scalar fields can be chosen later. The metric can be rewritten in the form
\begin{equation}\label{M2a}
ds^2 = R^2\left({c_2}\phi^4 dx\cdot dx + \phi^{-2} d\phi^I d\phi^I\right),
\end{equation}
where $\phi^I$ is an eight-component vector of length $\phi$. The $AdS_4 \times S^7$
solution of 11-dimensional supergravity, with $N$ units of flux through the seven-sphere,
relates the radius $R$ to the 11-dimensional Planck length $l_p$ by
\begin{equation} \label{R6}
R^6 = 32\pi^2 N l_p^6 .
\end{equation}

There is no world-volume gauge field in this case, so the $S_1$ term in the action is just
\begin{equation} \label{M2S1}
S_1 = -T_{M2} \int \sqrt{- \det \left( G_{\a\b}
\right) }\, d^3 \s ,
\end{equation}
where $G_{\a\b}$ is the pullback of the 11-dimensional spacetime metric. Since
$T_{M2} = (4\pi^2 l_p^3)^{-1} $,
\begin{equation}
T_{M2}R^3 = \sqrt{2N}/\pi.
\end{equation}
Choosing static gauge $x^\m(\s) = \d^\m_\a \s^\a$, as before, yields
\begin{equation}
S_1 = - T_{M2} \int \sqrt{ - \det \left( c_2 R^2\phi^4  \eta_{\m\n}
+(R/\phi)^2 \pa_\m \phi^I \pa_\n \phi^I \right) } \, d^3 x,
\end{equation}
which can be rewritten as
\begin{equation}
S_1 = - \frac{\sqrt{2N}}{\pi} c_2^{3/2}\int \phi^6 \sqrt{-\det \left(\eta_{\m\n}
+ \frac{\pa_\m \phi^I \pa_\n \phi^I}{ c_2 \phi^{6} } \right) } \, d^3 x.
\end{equation}

Now we turn to $S_2 = \mu_2 \int A_3$. Here $A_3$ is the pullback of the M-theory three-form,
whose field strength is denoted $F_4 = d A_3$. This field strength is proportional
(with a coefficient denoted $k_2$) to the volume form for $AdS_4$. Therefore
\begin{equation}
F_4 =  k_2 c_2^{3/2} R^4 \phi^5 d\phi \wedge dx^0 \wedge dx^1 \wedge dx^2
= \frac{1}{6} k_2 c_2^{3/2}  R^4 d (\phi^6 dx^0 \wedge dx^1 \wedge dx^2) .
\end{equation}
Thus, by the same reasoning as in the D3-brane analysis, we obtain
\begin{equation}
S_2 = \frac{\mu_2 k_2 c_2^{3/2} R^4}{6}  \int \phi^6 d^3 x .
\end{equation}

We can now determine $\mu_2 k_2$ by requiring that $\mu_2 \int_{S^7} \star F_4 = 2 \pi N$.
This is the statement that there are $N$ units of flux threading the $S^7$.
To avoid ambiguity (and errors) in relating the normalization of $F_4$ to its
Hodge dual, it is important that the Hodge dual operation
be applied to a dimensionless expression. Since $F_4$ is not dimensionless to begin with,
this requires the choice of a basic unit of length.
We find that to get the desired result, the appropriate choice for this unit of length
is $2\pi l_p $. It would be desirable to have a better
explanation for this rule, but this is the best we can do without discussing fermions
(aside from the physical argument that there should be no force on the brane).
When fermions are included, kappa symmetry will relate $S_2$ to $S_1$ and determine this normalization unambiguously. This issue did not
arise in the D3-brane problem, because $F_5$ was self-dual.

Supplying the appropriate power of $2\pi l_p$, we have
\begin{equation}
\star F_4 = \frac{ k_2 R^7}{(2\pi l_p)^3} {\rm vol}(S^7),
\end{equation}
where ${\rm vol}(S^7)$
denotes the volume form for a unit radius seven-sphere. Such a sphere has volume $\pi^4/3$, and thus
\begin{equation} \label{flux4}
\mu_2\int_{S^7}\star F_4 = \frac{\mu_2 k_2 R^7}{(2\pi l_p)^3}\frac{\pi^4}{3} =2\pi N .
\end{equation}
It follows that the coefficient in $S_2$ is
\begin{equation}
\frac{\mu_2 k_2 R^4}{6} = \frac{2\pi N (2\pi l_p)^3}{2\pi^4 R^3} = \frac{\sqrt{2N}}{\pi}.
\end{equation}
This is exactly what we want so that
\begin{equation}
S = S_1 + S_2 = - \frac{\sqrt{2N}}{\pi}c_2^{3/2} \int \phi^6 \left[\sqrt{-\det \left(\eta_{\m\n}
+ \frac{\pa_\m \phi^I \pa_\n \phi^I}{ c_2 \phi^{6} }\right) } -1 \right]\, d^3 x.
\end{equation}
As in the D3-brane problem, we find cancellation of  the nonderivative terms, which would
otherwise give a net force on the brane. As before, the sign
of $S_2$ would be reversed for an anti-brane. Then there would be a potential
$V \sim \phi^6$, which implies that the anti-brane is attracted to $\phi =0$,
the origin of the moduli space (and the horizon of the AdS space).

Finally, we can determine the parameter $c_2$ by requiring that the $\phi$ kinetic term is
normalized in some particular way. For example, requiring that it is canonically normalized gives
$c_2 = \pi^2/(2N)$. We will make a different choice later.

\subsection{The M2-brane in $\mathbf {AdS_4 \times S^7/Z_k}$}

The next case we wish to consider should correspond to the ABJM conformal field theory.
In particular, we wish to derive a candidate HEA for $U(2)_k \times U(2)_{-k}$
superconformal gauge theory on the Coulomb branch. This theory has ${\cal N}  =6 $
(3/4 maximal) supersymmetry (for $k >2$). There are two integers in the problem,
$N$ is the number of units of flux, as before. The second integer, $k$, controls
the levels of the Chern--Simons terms in the superconformal Chern--Simons theory.
The dual 11-dimensional geometry is closely
related to that discussed for $AdS_4 \times S^7$, which corresponds to $k=1$.
The only difference is that the seven-sphere is modded out by the discrete
group $\IZ_k$. So we need to figure out how to modify the previous analysis
to describe this situation. One thing is obvious: we should replace the eight
real scalar fields $\phi^I$ by four complex scalar fields $\Phi^A$:
\begin{equation}
\Phi^A = \phi^A + i \phi^{A+4} \quad A=1,2,3,4,
\end{equation}
so that
\begin{equation}
\Phi^2 = \sum_{A=1}^{4} |\Phi^A|^2 = \sum_{I=1}^8 (\phi^I)^2 = \phi^2.
\end{equation}
Only a $U(4)$ subgroup of $SO(8)$ will survive. Moreover, only the $SU(4)$ part of this is
an R-symmetry group belonging to the superconformal symmetry group $OSp(6|4)$.

One may be tempted to incorporate two $U(1)$ gauge fields into the description.
However, as discussed in Sect.~2.3 of the ABJM paper, there is a simpler option. One
can add a single periodic real scalar coordinate $\t$, with period $2\pi$, and a compensating
local symmetry. The metric in Eq.~(\ref{M2a}) is modified to become
\begin{equation}
ds^2 = R^2\left({c_2}\Phi^4 dx\cdot dx + \Phi^{-2} D\Phi \cdot D\overline{\Phi}\right),
\end{equation}
where
\begin{equation}
D\Phi^A = d\Phi^A +i k^{-1} \Phi^A d\t \quad \and
\quad D\overline\Phi_A = d\overline\Phi_A - i k^{-1} \overline\Phi_A d\t.
\end{equation}
Equivalently, defining a one-form $B = k^{-1} d\t$,
\begin{equation}
D\Phi^A = (d+ iB)\Phi^A  \quad \and
\quad D\overline\Phi_A = (d-iB)\overline\Phi_A .
\end{equation}
The radius is now given by
\begin{equation} \label{R6k}
R^6 = 32\pi^2 kN l_p^6 .
\end{equation}

This metric is invariant under the local symmetry
$\Phi^A \to \exp(i\theta) \Phi^A$ and $\t \to \t -k\theta$,
since these imply that $D\Phi^A \to \exp(i\theta) D\Phi^A$.
One could use this ``gauge freedom'' to set
$\t = 0$, but this would still leave transformations with $\theta$ an integer multiple of
$2\pi/k$. In other words, one is left with the equivalence
\begin{equation}
\Phi^A  \sim e^{2\pi i /k}\Phi^A.
\end{equation}
Thus, the resulting geometry is the desired $AdS_4 \times S^7/Z_k$. The following discussion
uses the gauge invariant formulation.
Equivalently, we can set $\t = k \s$ and $B_\m = \pa_\m \s$, provided we enforce a $2\pi/k$ periodicity in $\s$ by means of a suitable Chern--Simons term. The only other significant
modification of previous section concerns the flux quantization condition,
which now becomes $\mu_2 \int_{S^7/\IZ_k} \star F_4 = 2 \pi N$. This is accounted for
by replacing $N$ by $kN$ in Eqs.~(\ref{R6}) and (\ref{flux4}). This results in
\begin{equation}
S = S_1 + S_2 = - \frac{\sqrt{2kN}}{\pi}c_2^{3/2} \int \Phi^6 \left[\sqrt{-\det \left(\eta_{\m\n}
+ \frac{{\rm Re}\left[D_\m \Phi^A {D_\n \overline\Phi_A} \right]}{ c_2 \Phi^{6} }\right) } -1 \right]\, d^3 x.
\end{equation}
The kinetic term for the scalar fields is then
\begin{equation}
S_{\rm kin} = -\frac{k}{\pi}\sqrt{\frac{c_2 \l}{2}}\int {\rm Re}\left[D_\m \Phi^A
{D_\n \overline\Phi_A} \right] \, d^3 x,
\end{equation}
where we have introduced the 't Hooft parameter
\begin{equation}
\l = N/k.
\end{equation}

The quantity $D\Phi^A D\overline\Phi_A$ can be recast as follows
\begin{equation}
D\Phi^A D\overline\Phi_A = d\Phi^A d\overline\Phi_A +2 \Phi^2 B W +\Phi^2 B^2
= d\Phi^A d\overline\Phi_A - \Phi^2 W^2 +\Phi^2 (B+W)^2,
\end{equation}
where
\begin{equation}\label{Wdef}
W = \Phi^{-2} {\rm Im}\left[ \Phi^A d \overline\Phi_A\right].
\end{equation}
These formulas enable us to recast the determinant in the action in the form
\begin{equation}
\D = \det\left(G_{\m\n} + \frac{(B_\m + W_\m)(B_\n + W_\n)}
{c_2 \Phi^4}\right),
\end{equation}
where
\begin{equation}\label{GM2}
G_{\m\n} = \eta_{\m\n} + \frac{{\rm Re}\left[\pa_\m \Phi^A
{\pa_\n \overline\Phi_A} \right]}{ c_2 \Phi^{6} } - \frac{ {\rm Im}\left[ \Phi^A \pa_\m \overline\Phi_A\right]  {\rm Im}\left[ \Phi^A \pa_\n \overline\Phi_A\right]}{c_2 \Phi^8}.
\end{equation}
Because the second term inside the determinant is now rank one, we can recast the
square root of the determinant in the form
\begin{equation}
\sqrt{-\D} = \sqrt{-G} \sqrt{1 + \frac{(B+W)^2}{c_2 \Phi^4}},
\end{equation}
where $(B+W)^2 = G^{\m\n}(B_\m + W_\m)(B_\n + W_\n)$. (Inner products in this
paper use the Lorentz metric unless otherwise specified.) This has successfully isolated
the $B$ dependence, which was the purpose of these maneuvers. The action is now
\begin{equation}
S =  - \frac{k}{\pi}\sqrt{2\l} c_2^{3/2} \int \Phi^6 \left[ \sqrt{-G}
\sqrt{1 + \frac{(B+W)^2}{c_2 \Phi^4}}-1 \right]\, d^3 x.
\end{equation}

Next, let us consider treating $B$ as an
independent field and adding a Lagrange multiplier term to $S$:
\begin{equation}
S' = \frac{k}{4\pi}\int F \wedge (B - d \s)
= \frac{k}{4\pi}\int \e^{\m\n\r} F_{\m\n} (B_\r - \pa_\r \s) d^3 x
\end{equation}
Solving the equation of motion for the Lagrange multiplier $F$ restores the
original action. On the other hand,
the $\s$ equation of motion implies that $dF =0$ so that we can write $F=dA$.
The $B$ equation of motion now becomes
\begin{equation}\label{Beom}
\frac{k}{4\pi}\e^{\m\n\r} F_{\m\n} = \frac{k}{\pi}\sqrt{2c_2\l} \sqrt{-G}
\frac{(B^\r + W^\r)\Phi^2}{\sqrt{1 + \frac{(B+W)^2}{c_2 \Phi^4}}}.
\end{equation}
Squaring both sides one finds that
\begin{equation}
\left(1 + \frac{F^2}{16\l c_2^2\Phi^8}\right) \left(1 + \frac{(B+W)^2}{c_2\Phi^4}\right) =1.
\end{equation}
This allows one to solve the previous equation for $B_\m + W_\m$.
A nice choice of normalization is $c_2 = (8\l)^{-1}$, which gives
\begin{equation}
\left(1 + 4 \l \Phi^{-8}{F^2}\right) \left(1 + 8\l \Phi^{-4}{(B+W)^2}\right) =1.
\end{equation}
We will compare these equations to ones for a D2-brane probe in the next subsection.
The transformation we have made equates the Bianchi identity for $F$ with the
equation of motion for $B$. Thus, the vanishing of the divergence of the right-hand side
of Eq.~(\ref{Beom}) is the latter equation.

\subsection{The D2-brane in ${\mathbf{AdS_4 \times CP^3}}$}

Following ABJM \cite{Aharony:2008ug}, it is instructive to consider
a type IIA superstring theory background that corresponds to a certain limit
of the M-theory one. String theory in this background has a string coupling constant
$g_s$ and a perturbative string expansion. The appropriate background geometry,
$AdS_4 \times CP^3$,  has $OSp(6|4)$ superconformal
symmetry once the fermionic degrees of freedom are included. The bosonic truncation,
which is all that will be considered here, only
exhibits the $Sp(4)$ conformal symmetry and the $SU(4)$ R symmetry.

We need to consider a 2-brane to obtain
a three-dimensional conformal field theory, and the only BPS 2-brane in type IIA
superstring theory is the D2-brane. This immediately leads to a surprising conclusion.
The conformally invariant effective action that will result from treating this
D2-brane by the methods of this paper will lead to the inclusion of a dynamical $U(1)$
gauge field. One usually argues that this is not possible: the kinetic term of such
a field is dimension four, but all terms must have dimension three. Certainly, there
are no dynamical gauge fields to be found in ABJM theory. The resolution of this paradox, which
will be exhibited by formulas shortly, is that the kinetic term of the gauge field
will have the form $ F^2/\Phi^2$, which does have dimension three. This is sensible,
because we are considering the theory on the Coulomb branch.

The metric for this problem is
\begin{equation} \label{D2metric}
ds^2 = R^2\left({c_2}\Phi^4 dx \cdot dx + \Phi^{-2} d\Phi^2 + ds^2_{CP^3}\right)),
\end{equation}
where we have copied the $AdS_4$ expression from Eq.~(\ref{M2metric}). The
radius $R$ is the same as in Eq.~(\ref{R6k}). However, the equality of the M2-brane and D2-brane
tensions implies that $l_p^3 = g_s l_s^3$. Therefore, expressed in string units,
the radius is given by
\begin{equation}
R^6 = 32\pi^2 g_s^2 kN l_s^6.
\end{equation}
Furthermore, since the radius of the M-theory circle is $R/k$, which in string units
must correspond to $g_s l_s$, one deduces that
\begin{equation} \label{R2A}
R^2 = 2^{5/2}\pi \sqrt{\l} l_s^2
\end{equation}
and
\begin{equation}
g_s = \sqrt{\pi}  (2\l)^{5/4}/N.
\end{equation}
As before, $\l = N/k$ is the 't Hooft coupling.

The $CP^3$ metric can be written in terms of homogeneous coordinates, exhibiting its
$SU(4)$ symmetry, as follows
\begin{equation}
ds^2_{CP^3} = \frac{dz^A d\bar z_A}{|z|^2} - \frac{|\bar z_A dz^A|^2}{|z|^4},
\end{equation}
where $|z|^2 = z^A \bar z_A$. This formula depends on four complex coordinates, but it describes
a six-dimensional manifold. The reason this works is that the formula has a local symmetry
under $z^A \to \l z^A$, where $\l$ is any nonzero complex function.

In the previous problems we were able to combine the $ \phi^{-2} d\phi^2$ term in the $AdS$
metric with the metric of the compact space in a convenient manner. That is also the
case here. The key step is to use the local symmetry to set $|z| =\phi$. Doing this, and
renaming the coordinate $\Phi^A$, gives
\begin{equation}
ds_7^2 =\frac{(d\Phi)^2}{\Phi^2} + ds^2_{CP^3} =  \frac{d\Phi^2}{\Phi^2}
+ \frac{d\Phi^A d \overline\Phi_A}{\Phi^2} -
\frac{|\overline\Phi_A d\Phi^A|^2}{\Phi^4} = \frac{d\Phi^A d\overline\Phi_A}{\Phi^2}
- \frac{\left[ {\rm Im} (\overline\Phi_A d\Phi^A)\right]^2}{\Phi^4}.
\end{equation}
This formula still has a local $U(1)$ symmetry given by $\Phi^A \to e^{i\a} \Phi^A$. Therefore
it describes a seven-dimensional manifold. Since it incorporates the radial $AdS$
coordinate, it is obviously noncompact. Altogether, we have the ten-dimensional metric
\begin{equation}
ds^2 = R^2\left({c_2}\Phi^4 dx \cdot dx + ds^2_7\right)
= c_2 R^2 \Phi^4\left( dx \cdot dx + \frac{ds^2_7}{c_2\Phi^4}\right).
\end{equation}
The pullback of this metric to the three-dimensional D2-brane world volume is denoted
$c_2 R^2 \Phi^4 G_{\m\n}$. The crucial fact is that the $G_{\m\n}$ obtained here,
evaluated in static gauge, is precisely the same expression obtained in Eq.~(\ref{GM2}).

We can now write down the D2-brane action. Using the fact that the $S_2$ cancels the
potential term as in the previous examples, it is in static gauge
$$
S = -T_{D2} \int \left(\sqrt{- \det (c_2 R^2 \Phi^4 G_{\m\n} + 2\pi \a' F_{\m\n})}
- (c_2 R^2 \Phi^4)^{3/2}\right)\, d^3 \s
$$
\begin{equation} \label{D2S1}
= -\b \int \Phi^6 \left(\sqrt{- \det ( G_{\m\n} + \g \Phi^{-4} F_{\m\n})}-1\right)\, d^3 \s,
\end{equation}
where
\begin{equation}
\b = T_{D2} R^3 c_2^{3/2} = \frac{\sqrt{2kN}}{\pi} c_2^{3/2},
\end{equation}
\begin{equation}
\g = \frac{2\pi\a'}{c_2R^2} = (2 c_2 \sqrt{2\l})^{-1}.
\end{equation}
Choosing static gauge, as before, and expanding the action to extract the kinetic terms
for the scalar fields $\Phi^A$ and the $U(1)$ gauge field, one finds
\begin{equation}
-\frac{\b}{2c_2} \int {\rm Re}\left[\pa^\m \Phi^A \pa_\m \overline\Phi_A\right] d^3x
- \frac{1}{4} \b \g^2 \int \Phi^{-2} F^{\m\n} F_{\m\n} d^3 x.
\end{equation}
The only freedom in these terms concerns the $\Phi$ normalization which is encoded in the $c_2$
dependence. If one chooses
\begin{equation}
c_2 = (8\l)^{-1}
\end{equation}
then the kinetic terms become
\begin{equation}
-\frac{k}{4\pi} \int \pa^\m {\rm Re}\left[\Phi^A \pa_\m \overline\Phi_A\right] d^3x
- \frac{k}{8\pi} \int \Phi^{-2} F^{\m\n} F_{\m\n} d^3 x.
\end{equation}

The determinant in the action can be evaluated by methods similar to those of the previous
section. One obtains
\begin{equation}
\sqrt{- \det ( G_{\m\n} + \g \Phi^{-4} F_{\m\n})}
= \sqrt{-G} \sqrt{1 + \half\g^2 \Phi^{-8}F\cdot F },
\end{equation}
where $F\cdot F = G^{\m\r} G^{\n\l} F_{\m\n}F_{\r\l}$.

Let us now treat $F$ as an independent field and add a Lagrange multiplier term to ensure that $F=dA$ in analogy with the discussion in the previous section. Then the action becomes
\begin{equation}
-\b \int \Phi^6 \left(\sqrt{-G} \sqrt{1 + \half\g^2 \Phi^{-8}F\cdot F }-1\right)\, d^3 x
+ \frac{k}{4\pi} \int [B \wedge (F - dA) +W \wedge F]
\end{equation}
The $B$ equation gives $F=dA$ and the $A$ equation gives $dB=0$, which allows us to set
$B=d\s$. The $WF$ term must be part of $S_2$ that we have omitted until now. It is required
in order that the $F$ variation produces the combination $B+W$, which is required
for consistency with the previous section. In fact, ABJM point out that the ten-dimensional background contains a RR two-form field strength $F_2 \sim k J$, where $J = dW$. So this
term is actually a known part of $S_2$.

The $F$ equation of motion is
\begin{equation}
\frac{k}{8\pi\sqrt{-G}} \e^{\m\n\r} (B_\m  +W_\m)
= \half \frac{\b\g^2\Phi^{-2} F^{\n\r}}{\sqrt{1 + \half\g^2 \Phi^{-8}F\cdot F }}.
\end{equation}
Squaring this gives
\begin{equation}
(1 + \half \g^2 \Phi^{-8} F\cdot F )(1 + c_2^{-1} \Phi^{-4}(B+W)^2) =1.
\end{equation}
For the nice normalization choice $c_2  =(8\l)^{-1}$ this becomes
\begin{equation}
(1 + 4\l \Phi^{-8} F\cdot F )(1 + 8\l \Phi^{-4}(B+W)^2) =1
\end{equation}
in perfect agreement with what we found in the previous subsection.

\subsection{Summary}

To summarize, we have found two equivalent formulations of the ABJM probe-brane action that
are related by a duality transformation.\footnote{This is analogous to what was demonstrated for
flat spacetime backgrounds in \cite{Townsend:1995kk}\cite{Witten:1995ex}.}
The D2-brane formulation has the field content of an abelian super Yang--Mills theory, and
the M2-brane formulation has the field content of an abelian superconformal Chern--Simons
theory. The only information that is lacking in the D2-brane derivation, by itself,
is the fact that the string coupling constant should take the form
$g_s = \sqrt{\pi} (2N/k)^{5/4}/N$, where $k$ and $N$ are positive integers.

For the preferred normalization choice $c_2  =(8\l)^{-1}$ the
complete D2-brane action is
\begin{equation} \label{D2final}
S = -\frac{k}{16\pi\l}\int \Phi^6 \left(\sqrt{- \det ( G_{\m\n}
+ \sqrt{8\l} \Phi^{-4} F_{\m\n})}-1\right)\, d^3 \s
+ \frac{k}{4\pi} \int W \wedge F,
\end{equation}
where $F=dA$, and $W$ and $G_{\m\n}$ are defined in Eqs.~(\ref{Wdef}) and (\ref{GM2}).
Since $k \sim \l^{1/4}/g_s$,
we are finding that the action is $1/g_s$
times an expression that only involves $\l$, just as we found for the D3-brane problem.
As before, this suggests that the loop expansion of the HEA corresponds to the topological
expansion of the nonabelian theory. We have conjectured that the case of relevance
to the $U(2)_k \times U(2)_{-k}$ HEA is $N=1$, which implies $\l = 1/k$.
The dual M2-brane action, for the same choice of normalization, is
\begin{equation}
S =  - \frac{k}{16\pi\l}\int \Phi^6 \left[\sqrt{-\det \left(\eta_{\m\n}
+ 8\l\frac{{\rm Re}\left[D_\m \Phi^A {D_\n \overline\Phi_A} \right]}{\Phi^{6} }\right) }
-1 \right]\, d^3 x
+\frac{k}{4\pi} \int A \wedge dB,
\end{equation}
where $B$ is the connection used in the definition of the covariant derivatives.
By defining $A^{\pm} = (A \pm B)/2$, the last term can be reexpressed as the
sum of the two $U(1)$ Chern--Simons terms with levels $k$ and $-k$.

One interesting problem is the study of soliton solutions. We expect to find
a soliton solution that can be interpreted as the fundamental
string ending on the D2-brane. This should describe the massive fields of the ABJM theory
that have been integrated out. In addition, there should be an instanton that can
be interpreted (in the Euclideanized theory) as a D0-brane ending on the D2-brane. It
is analogous to the monopole of the four-dimensional theory, which corresponds to
a D1-brane ending on a D3-brane.

\section{The M5-brane in $\mathbf {AdS_7 \times S^4}$}

In this case the $AdS_7$ radius is twice the $S^4$ radius $R$. So the
11-dimensional metric is
\begin{equation}\label{M5metric}
ds^2 = (2R)^2 \left( v^2 dx^2 + v^{-2} dv^2\right) + R^2 d\O_4^2
= R^2\left(c_5\phi dx^2 + \phi^{-2} d\phi^2 + d\O_4^2\right).
\end{equation}
Here we have made the change of variables $v = \sqrt{c_5\phi}/2$. This is motivated by
the fact that $v$ has dimension 1, whereas a scalar field in six dimensions
should have dimension 2. This can be rewritten in the form
\begin{equation}
ds^2 = R^2\left(c_5\phi dx^2 + \phi^{-2} d\phi^I d\phi^I\right).
\end{equation}
Here $\phi^I$ is a five-component vector of length $\phi$. The radius $R$ is given
in terms of the 11-dimensional Planck length $l_p$ by
\begin{equation}
R^3 = \pi N l_p^3.
\end{equation}
Therefore, we have (in static gauge)
\begin{equation}\label{M5S1}
S_1 = -T_{M5} R^6 c_5^3\int \phi^3 \sqrt{-\det\left(\eta_{\m\n}
+ \frac{\pa_\m \phi^I \pa_\n \phi^I}{c_5\phi^3} + i\tilde H_{\m\n}\right) }\, d^6 x.
\end{equation}
Since $T_{M5} = 2\pi/(2\pi l_p)^6$, the coefficient of $S_1$ simplifies to
\begin{equation}
T_{M5} R^6 = \frac{N^2}{32\pi^3}.
\end{equation}

The $\tilde H$ term encodes information about the world-volume field that is a two-form
with a self-dual three-form field strength. The knowledge of how it does that is not
required to understand the discussion that follows. However, here is a very brief
description of the method for doing this that was developed in \cite{Perry:1996mk}
and \cite{Schwarz:1997mc} and applied to the M5-brane problem in
\cite{Pasti:1997gx} and \cite{Aganagic:1997zq}. The procedure entails treating one
spatial dimension differently from the other five and making general coordinate
invariance manifest for only five of the six dimensions. The symmetry
in the sixth dimension is also there, but it is implemented in a more complicated,
less manifest, way. First one defines the five-dimensional restriction of the
the six-dimensional three-form $H =dB$, which in components is $H_{\m\n\r}$.
Then one defines the five-dimensional dual $\tilde H^{\m\n}$, and lowers the
indices with a six-by-five piece of the six-by-six metric tensor to give
$H_{\hat\m \hat\n}$, where hatted indices are six dimensional. The matrix
appearing in Eq.~(\ref{M5S1}) is six by six; the hats have been omitted.

Our immediate objective is to check whether $S_2$ cancels the leading nonderivative term,
as in the previous examples.
We want $S_2 = \m_5 \int A_6$, but what is the dual potential $A_6$?
In general, the equations of motion of M-theory give a rather complicated expression
for $d\star F_4$, including
a term proportional to $F_4 \wedge F_4$ among others. However, in the special case
of the $AdS_7 \times S^4$ solution, which is known to be an exact solution of M-theory,
one has $d\star F_4=0$.
Therefore in this background it is correct to define the dual potential by $dA_6 = \star F_4$.

By the same reasoning as in
the previous examples, we write $F_4 = k_5 R^4 {\rm vol} (S^4)$,
where ${\rm vol} (S^4)$ is the volume four-form for a four-sphere of unit radius.
Then, referring to the metric in Eq.~(\ref{M5metric}) and inserting a power of $2\pi l_p$ as before,
we see that
\begin{equation}
\star F_4 =  \frac{k_5  c_5^3 R^7}{(2\pi l_p)^3} \phi^2 d\phi \wedge dx^0 \wedge \ldots \wedge dx^5
= \frac{1}{3}\frac{k_5  c_5^3 R^7}{(2\pi l_p)^3} d (\phi^3 dx^0 \wedge \ldots \wedge dx^5) .
\end{equation}
and hence
\begin{equation}
S_2 = \frac{1}{3}\frac{k_5 \m_5 c_5^3 R^7}{(2\pi l_p)^3} \int \phi^3 d^6 x.
\end{equation}
Next we require that
\begin{equation}
\m_5 \int F_4 = \mu_5 k_5 R^4 \int_{S^4} {\rm vol} (S^4) = \m_5 k_5 R^4( 8 \pi^2 /3) = 2\pi N.
\end{equation}
This gives
\begin{equation}
\frac{1}{3}\frac{k_5 \m_5 R^7}{(2\pi l_p)^3} = \frac{N^2}{32\pi^3}.
\end{equation}

We conclude that
\begin{equation}
S = S_1 + S_2= - \frac{N^2 c_5^3}{32\pi^3}\int \phi^3 \left[\sqrt{-\det\left(\eta_{\m\n}
+  \frac{\pa_\m \phi^I \pa_\n \phi^I}{c_5\phi^3} + i\tilde H_{\m\n}\right) }-1\right]\, d^6 x.
\end{equation}
Once again, we find the desired cancellation of the potential.

One of the important motivations for this work was the desire to formulate
effective actions for $(2,0)$ theories in six dimensions on the Coulomb branch.
The supersymmetric completion of the bosonic
M5-brane action, given above, is a good candidate
for the simplest such theory. Its applications are limited, however,
because it contains no parameter on which to base
a perturbation expansion. However, there should be controlled
expansions, at energies small compared to the scale set by the vev
of the scalar field. This might justify looking for soliton solutions
of the classical action. This would
be somewhat analogous to looking for soliton solutions of eleven-dimensional
supergravity. The most interesting soliton candidate is a self-dual
string. Such a solution has already been found for the M5-brane probe action in
flat 11-dimensional spacetime \cite{Howe:1997ue}.

In the M5-brane case there is no known analog of ABJM orbifolding, which is the
way a perturbative expansion was made possible in the case of the M2-brane theory.
Despite its limitations, it is intriguing that there is a candidate for the
six-dimensional $(2,0)$ HEA in the Coulomb phase.
After all, there is no known Lagrangian formulation of the unbroken phase.
In fact, is generally assumed that such a formula does not exist.
That makes this action all the more significant.
The utility of the M5-brane formula may improve
after compactification, although this breaks the conformal symmetry explicitly.

\section{Conclusion}

We have conjectured that the world-volume action of a probe $p$-brane in a maximally
supersymmetric spacetime of the form $AdS_{p+2} \times S^n$ (or 3/4 maximally
supersymmetric in the case of $AdS_4 \times CP^3$) can be reinterpreted as the
solution to a different problem: finding an explicit formula for the {\em highly effective
action (HEA) of a superconformal field theory in $p+1$ dimensions on the Coulomb branch}.
The bosonic truncations of a few such probe $p$-brane
world-volume theories were described in detail. The main evidence in
support of the conjecture is that the actions incorporate all of the expected symmetries
and dualities. In the case of the D3-brane these include $PSU(2,2|4)$ superconformal
symmetry (when fermions are included)
and $SL(2,\IZ)$ duality. The methodology of the constructions ensures that
these properties are built into the formulas. Nevertheless,
S-duality was verified explicitly in Appendix B.

The actions derived by considering brane probes possess
local symmetries: general coordinate invariance and
local kappa symmetry. This fact seems quite profound, and it could be
an important clue to possible generalizations. We described a specific gauge choice
that gives an action with manifest Poincar\'e and scaling symmetry
and the expected field content. It will be accompanied by a fermionic gauge choice
for the complete theory with local kappa symmetry. However,
it seems reasonable to regard the HEAs with the local
symmetries as more fundamental. After all, that is a useful point of view for
other gauge theories such as Yang--Mills theory and general relativity.

Important projects for the future are the incorporation of
fermions and comparisons of expansions of the resulting formulas with existing
results for Coulomb-branch low-energy effective actions in the literature.
One feature of HEAs that should be tested is the fact that
the probe-brane approach naturally leads to formulas in which only first
derivatives of fields appear.
Is there some deep reason why there are no higher derivatives? To some extent this is a
formalism-dependent question. Some classes of higher-derivative
terms can be introduced or removed by field redefinitions. It may be worthwhile to
explore the extent to which derivative terms can be removed in this way. Also,
derivatives can be moved around by adding total derivative terms to the action.
These comparisons will require choosing static gauge, since that is all that
standard approaches to the construction of low-energy effective actions can reproduce.
One can also explore whether agreement with existing results require the choice $N=1$.

It should be interesting to construct soliton and instanton solutions of the
classical HEAs. In the case of the D3-brane theory, for example, we expect to find a
complete $SL(2,\IZ)$ multiplet of solitons containing $W$, monopole, and dyon supermultiplets.
Much more challenging projects will be the construction of
multi-soliton and multi-instanton solutions and the exploration of their moduli spaces.

The actions derived in this paper have all of the desired symmetry and duality properties
for any positive integer $N$ (the number of units of flux). We have proposed
that the correct choice is $N=1$, even though that is the choice
for which the probe approximation is most suspect.
We don't understand well the role of the $N>1$ actions from the point of
view of effective actions. However, if we take the $N$ dependence at face value,
the structure of the formulas suggests that the loop expansion of the
D-brane HEAs should correspond
to the string loop expansion or equivalently the topological expansion of
the Coulomb-branch gauge theory with explicit $W$ fields.
Given our uncertainty about the meaning of $N$,
this conjecture could be false even if we have
correctly identified the $N=1$ theory as the HEA. However, if this additional
conjecture is correct, it would imply that the classical HEA action
and the tree approximation to scattering amplitudes should exhibit dual
conformal symmetry in addition to conformal symmetry, and hence have the
full Yangian symmetry.

When one incorporates fermions into the brane-probe actions, one will need
to decide whether to use component fields or superfields. The formulas
will probably be derived first in terms of component fields, but it should
be useful for some purposes to recast them in terms of superfields.
One could try to do this by modifying the
probe-brane analysis or by trying to supersymmetrize the bosonic truncation.
There has been some interesting work on superfield formulations
of supersymmetric extensions of Born--Infeld theory that retain duality symmetry
\cite{Bagger:1996wp} -- \cite{Kuzenko:2004sv}. Some of these papers emphasize
the relationship between brane-probe actions and Coulomb-branch {\em low-energy}
effective actions.

If our conjecture survives further scrutiny, it will become important to understand
the extent to which symmetry and other general considerations determine the
HEA. One will also want to understand {\em why} the world-volume theory
of a brane probe should give an HEA. Clearly, the two problems are
not completely unrelated. After all, the brane probe provides information about
a $U(N+1)$ theory that is broken to $U(N) \times U(1)$ with
the brane probe most closely related to the $U(1)$ factor.
However, that interpretation seems to require
a large-$N$ approximation. In any case, we have made a precise conjecture for the
$N=1$ brane-probe action, but not for those with $N>1$.

Another (possibly related) problem concerns the construction of effective actions for
higher-rank gauge theories on the Coulomb branch. When the gauge theory has rank $r>1$,
the HEA should contain $r$ massless abelian supermultiplets. It is not
clear how to generalize the brane constructions discussed here to address such
cases. Presumably, the formulas obtained here with $N>1$ are ingredients in
these constructions, but it seems likely that additional,
more complicated, ingredients are also required.

The analysis in this paper only treated HEAs classically, though
this already encodes a lot of quantum information. There is still
much to be learned about the classical theories (inclusion of fermions, structure
of tree amplitudes, solitons, possible Yangian symmetry, etc.), so it may be
premature to explore their quantization.

In conclusion, we have described a few specific cases in which
the world-volume action of a probe brane in an anti de Sitter background with
one unit of flux is a good candidate for the
highly effective action of a superconformal field theory
on the Coulomb branch. Furthermore, in the case of the ${\cal N}=4$
and ABJM examples, it seems likely that the
loop expansion of the HEA corresponds to the
topological expansion of the nonabelian gauge theory with explicit $W$ fields
on the Coulomb branch.
It will be exciting to see whether these conjectures survive further scrutiny.

\noindent{\em Notes added:}

A. Tseytlin has pointed out that
there is evidence of a disagreement between the D3-brane probe action and
the ${\cal N}=4$ SYM Coulomb-branch effective action, which is described
in \cite{Buchbinder:2001ui}. He claims that
the disagreement, which concerns the one-loop $F^8$ terms, is not
removable by a field redefinition. The author is grateful to Tseytlin
for bringing this to his attention.

Another proposal for a relationship between gauge theories
and probe D-brane actions, which seems to be very different from the
one discussed in this paper, has appeared recently \cite{Ferrari:2013aba}. It would be
interesting to explore how the two proposals are related.

\section*{Acknowledgment}

The author wishes to thank Savdeep Sethi for suggesting that he investigate effective
actions for $(2,0)$ superconformal field theories on the Coulomb
branch and for drawing \cite{Sethi:1996kj}\cite{Maxfield:2012aw} to his attention.
He also acknowledges discussions with
Hee-Joong Chung, Sergei Gukov, Nicholas Hunter-Jones, Arthur Lipstein, and Wenbin Yan.
This work was supported in part by the U.S. Dept. of Energy
Grant No. DE-FG03-92-ER40701.  The author acknowledges the hospitality of the Aspen Center
for Physics, where he began working on this project. The ACP is supported by the National
Science Foundation Grant No. PHY-1066293.

\newpage

\appendix

\section{The AdS Poincar\'e patch}

Consider the description of $AdS_{p+2}$ with unit radius given by the hypersurface
\begin{equation}\label{hyper}
y \cdot y -uv = -1,
\end{equation}
where $ y \cdot y = -(y^0)^2 + \sum_1^p (y^i)^2$ is a Lorentzian product in $p +1$ dimensions.
The Poincar\'e-patch metric for $AdS_{p+2}$ with radius $R$ is then
\begin{equation}
ds^2 = R^2 (dy\cdot dy - du dv).
\end{equation}
Defining $x^\m = y^\m/v$ and eliminating $u = v^{-1} + v x \cdot x$ then gives
\begin{equation} \label{AdSmetric}
ds^2 = R^2(v^2 dx\cdot dx + v^{-2}dv^2).
\end{equation}
The coordinate $x^\m$ has dimensions of length, and $v$ has the dimensions of inverse length,
as is appropriate for a scalar field in four dimensions.

The hypersurface in Eq.~(\ref{hyper}) has $SO(p+1,2)$ symmetry, which corresponds
to the conformal symmetry of a $p$-brane world-volume theory.
Poincar\'e invariance in $p+1$ dimensions and the scaling symmetry $x^\m \to \l x^\m$,
$v \to \l^{-1}v$ are manifest symmetries of the metric. The only symmetries that are not manifest
in the metric of Eq.~(\ref{AdSmetric}) are those that correspond to conformal transformations.
Infinitesimally, these are given by the hypersurface symmetries $\d y^\m =  b^\m u$,
$\d v = 2b\cdot y$, $\d u =0$. This corresponds to
\begin{equation}\label{conformal1}
\d x^\m =  b^\m (v^{-2} + x\cdot x) - 2 b \cdot x x^\m,
\quad \d v =2 b \cdot x v .
\end{equation}

In order to check the conformal symmetry of the ${AdS}$ metric, let us
rewrite Eq.~(\ref{AdSmetric}) in the form
\begin{equation}
ds^2 = R^2(\eta_{\m\n} e^\m e^\n + f^2),
\end{equation}
where
\begin{equation}
e^\m = vdx^\m \quad {\rm and} \quad f = v^{-1} dv.
\end{equation}
In terms of these one-forms the conformal transformations become
\begin{equation} \label{deltae}
\d e^\m = -2b^\m v^{-1}  f + 2b^\m x \cdot e - 2x^\m b \cdot e
\end{equation}
\begin{equation}\label{deltaf}
\d f =  2v^{-1} b \cdot e .
\end{equation}
Using these it is easy to verify that $e \cdot \d e +  f  \d f =0$,
which proves that the metric is invariant.

\section{S-duality of the D3-brane action}

As shown in Eq.~(\ref{taudef}), the complex parameter $\t = \t_1 + i \t_2$ is the background
value of the complex scalar field of the type IIB supergravity multiplet.
It transforms under the $SL(2,\IZ)$ S-duality group in the usual nonlinear fashion,
$\t \to (a \t + b)(c\t + d)^{-1}$, where $a,b,c,d$ are integers satisfying $ad-bc=1$.
This symmetry of the D3-brane action is induced from the $SL(2,\IZ)$ symmetry of
type IIB superstring theory. We will focus our attention on the $S$ transformation $\t \to \t' = -1/\t$,
or
\begin{equation}
\t_1 \to \t_1'= -\frac{\t_1}{|\t|^2} \quad {\rm and} \quad
\t_2 \to \t_2' = \frac{\t_2}{|\t|^2}.
\end{equation}
Much of the discussion of duality in the literature is specific to the self-dual point $\t=i$.
At this point, the unbroken subgroup of the {\em classical} $SL(2, \IR)$ duality group is $SO(2)$.
S-duality then corresponds to a rotation by $\pi/2$, while a rotation by $\pi$ correspond to sending
the fields to their negatives. We will consider the case of arbitrary $\t$ (with $\t_2 >0$), and focus attention
on the $S$ transformation described above.

Let us begin with free Maxwell theory in a background metric $G_{\m\n}$ written in the form
\begin{equation} \label{freeMaxwell}
S = - \frac{\t_2}{8\pi} \int F\cdot F \sqrt{-G} d^4x + \frac{\t_1}{8\pi} \int F \wedge F
= \frac{1}{8\pi} \int (\t_1 F \cdot \tilde F - \t_2 F\cdot F  ) \sqrt{-G} d^4 x,
\end{equation}
where $G_{\m\n}$ is a Lorentzian-signature metric tensor and  $G$ is its determinant.
$F= dA$ is the usual two-form field strength constructed from a one-form potential.
Also,
\begin{equation}
F\cdot F = G^{\m\rho} G^{\n \l} F_{\m\n} F_{\rho\l}
\end{equation}
and $\tilde F$ is the Hodge dual:
\begin{equation}
\tilde F^{\m\n} = \frac{\e^{\m\n\rho\l} F_{\rho\l}}{2\sqrt{-G}}.
\end{equation}
The $F \cdot \tilde F$ term is metric independent.

The Bianchi identity is $d F =0$, and the classical field equation is $ d \tilde F =0$.
The basic idea of electric-magnetic symmetry, or S-duality, is that the symmetry
$\t \to -1/\t$ can be understood by simultaneously passing to the dual potential,
thereby interchanging the roles of the Bianchi identity and the equation of motion.
To understand how this works, consider the action with an additional Lagrange multiplier term:
\begin{equation}
S = \frac{1}{8\pi} \int \left( \t_1 F \cdot \tilde F - \t_2 F\cdot F -2 \tilde H^{\m\n}
(F_{\m\n} - 2\pa_\m A_\n)  \right) \sqrt{-G} d^4 x.
\end{equation}
Here $F_{\m\n}$ is treated as an independent field. The equation
$F_{\m\n} = \pa_\m A_\n - \pa_\n A_\m$ arises as a consequence of solving the $H$
equation of motion. In this way one
returns to the original action.

The $A$ equation of motion is
solved by introducing a dual potential $A'_\m$ with $H_{\m\n} = \pa_\m A'_\n - \pa_\n A'_\n$.
Furthermore, the $F$ equation of motion gives
\begin{equation}
\tilde H_{\m\n} = \t_1 \tilde F_{\m\n} - \t_2 F_{\m\n}.
\end{equation}
The inversion of this formula is
\begin{equation}\label{freeinverse}
-\tilde F_{\m\n} = \t_1' \tilde H_{\m\n} - \t_2' H_{\m\n} .
\end{equation}
From this it follows that the action
\begin{equation}
S_{\rm dual} = \frac{1}{8\pi} \int \left( \t_1' H \cdot \tilde H - \t_2' H\cdot H
+2 \tilde F^{\m\n} (H_{\m\n} - 2\pa_\m A'_\n)  \right) \sqrt{-G} d^4 x,
\end{equation}
gives rise to exactly the same set of equations. This
proves that $S$ and $S_{\rm dual}$, without the Lagrange multiplier terms
(but with $F=dA$ and $H=dA'$), are equivalent. This proves S-duality for free Maxwell theory
in an arbitrary background geometry. Note that the minus sign on the left-hand side
of Eq.~(\ref{freeinverse})
has been accounted for by reversing the sign of the Lagrange multiplier term.
This analysis is valid at the quantum level. As evidence of this, note that
the topological term, which is proportional to
$\t_1 \int F \wedge F$, does not contribute to the classical equations of motion, but it
plays an important part in the analysis. The same will be true for the more
complicated formulas in the remainder of this appendix.

Let us turn next to the verification of S-duality for the D3-brane action. We should
simply substitute Eq.~(\ref{finalD3}) in place of Eq.~(\ref{freeMaxwell}) and repeat
the same steps. The analysis that follows is based on methods introduced in
\cite{Tseytlin:1996it} -- \cite{Aganagic:1997zk}.
(For earlier work on this subject see
\cite{Gaillard:1981rj}, \cite{Zumino:1981pt}.)
The algebra is a bit overwhelming if one tackles the full problem head on.
So let us approach it in a few steps.
We begin with the somewhat simpler problem given by the action
\begin{equation}
S = \frac{1}{8\pi} \int \left(\t_1 F\cdot \tilde F  - 4 \D(F, \t_2)
-2 \tilde H^{\m\n} (F_{\m\n} -2\pa_\m A_\n)\right) d^4x,
\end{equation}
where
\begin{equation}
\D(F, \t_2) =  \sqrt{-\det(\eta_{\m\n} + \sqrt{\t_2}F_{\m\n})}  = \sqrt{1 +\half \t_2 F\cdot F
- \frac{1}{16} \t_2^2(F \cdot \tilde F)^2}.
\end{equation}
As before, the $H$ equation implies that $F_{\m\n} = \pa_\m A_\n - \pa_\n A_\m$,
and the $A$ equation is solved by writing $H_{\m\n} = \pa_\m A'_\n - \pa_\n A'_\m$.

The $F_{\m\n}$ field equation is
\begin{equation} \label{Heqn}
\tilde H^{\m\n} = R^{\m\n} (F, \t) \equiv \t_1 \tilde F^{\m\n} + \frac{\t_2 F^{\m\n}
-(1/4) {\t_2}^2 \tilde F^{\m\n} F \cdot \tilde F}{\D(F, \t_2)}.
\end{equation}
Then the remarkable theorem, which is the essential step in the proof of S-duality, is
that this equation can be solved for $F$ giving
\begin{equation}\label{inverse}
\tilde F^{\m\n} = -R^{\m\n} (H, \t') .
\end{equation}
Before discussing the proof of this formula, let us make clear what it implies. Just as
in the preceding discussion of the free theory, it implies that
\begin{equation}
S_{\rm dual} = \frac{1}{8\pi} \int \left(\t_1' H\cdot \tilde H  - 4 \D(H, \t_2') +2 \tilde F^{\m\n}
(H_{\m\n} - 2\pa_\m A'_\n)  \right)  d^4 x,
\end{equation}
gives exactly the same set of equations of as the original action. This
proves that $S$ and $S_{\rm dual}$, without the Lagrange multiplier terms
(but with $F=dA$ and $H=dA'$), are equivalent.

In principle, Eq.~(\ref{inverse}) can be verified by direct substitution, though this is
very difficult. The analysis can be simplified as follows: Without loss of generality,
the Lorentz covariance of the formulas allows one to choose a
Lorentz frame such that  $F$ takes the canonical form
\begin{equation}
F_{\m\n} = \begin{pmatrix} 0 & E & 0 & 0\\ -E & 0 & 0 & 0\\
0& 0 & 0& B\\  0& 0 & -B & 0 \end{pmatrix} .
\end{equation}
Then the only part of $H$ that contributes is
\begin{equation}
H_{\m\n} = \begin{pmatrix} 0 & E' & 0 & 0\\ -E' & 0 & 0 & 0\\
0& 0 & 0& B'\\  0& 0 & -B' & 0 \end{pmatrix} .
\end{equation}
The six components of Eq.~(\ref{Heqn}) then simplify to the two formulas
\begin{equation}
B' = \t_1 B + \t_2 E \sqrt{\frac{1+ \t_2 B^2}{1 - \t_2 E^2}}
\end{equation}
\begin{equation}
E' = \t_1 E - \t_2 B \sqrt{\frac{1 - \t_2 E^2}{1+ \t_2 B^2}}.
\end{equation}
Therefore the theorem we wish to verify simplifies to the statement that these two
equations for $E$ and $B$ are solved by
\begin{equation}
-B = \t_1' B' + \t_2' E' \sqrt{\frac{1+ \t_2' B'^2}{1 - \t_2' E'^2}}
\end{equation}
\begin{equation}
-E = \t_1' E' - \t_2' B' \sqrt{\frac{1 - \t_2' E'^2}{1+ \t_2' B'^2}}.
\end{equation}
With some effort, this can be verified by direct substitution. The special
case $\t_1 =\t_1'=0$ is easy to verify.

This is not yet the end. The next step is to introduce an arbitrary Lorentzian signature
background metric $G_{\m\n}$, as we discussed for the free theory. We first note that
\begin{equation}
\sqrt{-\det(G_{\m\n} + \sqrt{\t_2} F_{\m\n})} = \sqrt{-G} \D ,
\end{equation}
where
\begin{equation}
\D =  \sqrt{-\det(\eta_{mn} + \sqrt{\t_2}F_{mn})}  = \sqrt{1 +\half \t_2 F\cdot F
- \frac{1}{16} \t_2^2(F \cdot \tilde F)^2}.
\end{equation}
Here, we have implicitly introduced a vierbein $E^m_\m$, such that
$G_{\m\n} = E_\m^m \eta_{mn} E^n_\n$ and used its inverse to construct
$F_{mn} = E_m^\m E_n^\n F_{\m\n}$.  This means that the metric $G$ is used to define
index contractions and the Hodge dual in the final expression for $\D$. The previous
argument then goes through without any changes.

The action is now
\begin{equation}
S = \frac{1}{8\pi} \int  \left(\t_1 F\cdot \tilde F  - 4 \sqrt{-\det(\eta_{mn}
+ \sqrt{\t_2}F_{mn})} \right) \sqrt{-G} d^4x,
\end{equation}
but it is still not in the desired final form. The next step is to utilize the
freedom to redefine the metric by a Weyl transformation $G_{\m\n} \to f G_{\m\n}$,
which implies that $F_{mn} \to f^{-1} F_{mn}$. This transforms the action to
\begin{equation} \label{metricform}
S = \frac{1}{8\pi} \int  \left(\t_1 F\cdot \tilde F  - 4 f^2\sqrt{-\det(\eta_{mn}
+ f^{-1}\sqrt{\t_2}F_{mn})} \right) \sqrt{-G} d^4x
\end{equation}
By choosing $ f = {\sqrt{2\pi c_3} \phi^2} $, where $c_3 = \pi/\l$,
Eq.~(\ref{finalD3}) can be recast in the S-duality invariant form of Eq.~(\ref{metricform})
for the choice
\begin{equation}
G_{\m\n} = f\left( \eta_{\m\n} + \frac{\pa_\m \phi \cdot \pa_\n \phi}{c_3 \phi^4}\right)
= {\sqrt{2\pi c_3} \phi^2} \eta_{\m\n}
+ \frac{\sqrt{2\pi}}{\sqrt{c_3} \phi^2} \pa_\m \phi \cdot \pa_\n \phi.
\end{equation}

In this analysis of S-duality we have omitted the important force canceling term
proportional to $\int \phi^4 d^4x$ contributed by $S_2$. It does not depend on the
gauge field, and therefore
it does not effect the argument.

\section{The D3-brane in global $\mathbf {AdS_5 \times S^5}$}

The Poincar\'e patch description of AdS can be extended to a geodesically complete space. This
space has a periodic time coordinate $\theta$, but we can pass to the covering space by replacing
this circle by an infinite line described by the global time coordinate $t$. In this description
the $SO(2)$ subgroup (after the comma) of $SO(p+1,2)$ is replaced by its noncompact covering group
$\IR$. This the is translation symmetry group for the global time coordinate. Let's quickly review
the derivation of the formulas.

We first replace Eq.~(\ref{hyper}) for the hypersurface by the equivalent formula
\begin{equation} \label{hyper2}
\sum _{i=1}^{p+1} y_i^2 - t_1^2 - t_2^2 = -1.
\end{equation}
We then pass to spherical coordinates for the $y$'s and the $t$'s: $(y, \O_p)$ and $(\t, \th)$.
Replacing $\theta$ by $t$, the metric then takes the form
\begin{equation}
ds^2 =  R^2 \left(dy_i dy_i -\frac{y^2 dy^2}{1+y^2} - (1+y^2)dt^2\right)
= R^2 \left(\frac{dy^2}{1+y^2} + y^2 d\O_p^2 - (1+y^2)dt^2 \right).
\end{equation}
In these coordinates there is no horizon. The metric is nonsingular at $y=0$,
which is the center of global AdS. $y=\infty$ is the boundary of AdS. For the D3-brane
we simply add a five-sphere of radius $R$, so
\begin{equation}
ds^2  = R^2 \left(\frac{dy^2}{1+y^2} + y^2 d\O_3^2 - (1+y^2)dt^2 + d\O_5^2 \right).
\end{equation}


Making the change of coordinates $y =\half(\l^{-1} - \l)$ brings
the D3-brane metric to the form
\begin{equation}
ds^2 = R^2\left(\frac{ d\l \cdot d\l}{\l^2} + \frac{1}{4} (\l^{-1} - \l)^2 d\O_3^2
- \frac{1}{4} (\l^{-1} + \l)^2 dt^2\right).
\end{equation}
As in the discussion of the Poincar\'e patch description, $\l^I$ is a six-vector that
exhibits the $SO(6)$ symmetry of the five-sphere in a convenient way.
While these coordinates have some appeal,
we will work with $y$ rather than $\l$ in the remainder of this section.

Let us examine the world-sheet action, starting with $S_1$ as given in Eq.~(15). The counterpart
of Eq.~(18) is
\begin{equation}
S_1 = -\frac{N}{2\pi^2} \int  \sqrt{- \det \left(G_{\a\b}
+ \sqrt{{\pi}/(g_s N)}  F_{\a\b}\right)}\, d^4 \s,
\end{equation}
where $G_{\a\b}$ is the pullback of the ten-dimensional metric, with the factor $R^2$ removed:
\begin{equation}
 G_{\a\b} = \frac{\pa_\a y \pa_\b y}{1+y^2} + y^2 g_{ij} \pa_\a u^i \pa_\b u^j
 -(1+y^2) \pa_\a t \pa_\b t + h_{IJ} \pa_\a v^I \pa_\b v^J
\end{equation}
Here we have introduced coordinates for the spheres: $d\O_3^2 = g_{ij} du^i du^j$ and
$d\O_5^2 = h_{IJ} dv^I dv^J$. We do not need to be more explicit than that.

The static gauge choice in this case requires a world sheet of topology $S^3 \times \IR$. Then
one can identify the $S^3$ coordinates of the world sheet with those of the $S^3$ in the metric
and the time coordinate of the world sheet with the global time coordinate $t$.
In the static gauge the metric becomes $G_{\m\n} = K_{\m\n} + M_{\m\n}$, where
\begin{equation}
K_{\m\n} = \begin{pmatrix} - (1+y^2) & 0 \\ 0 &  y^2 g_{ij} \end{pmatrix}
\end{equation}
and
\begin{equation}
M_{\m\n} = \frac{\pa_\m y \pa_\n y}{1+y^2}  + h_{IJ} \pa_\m v^I \pa_\n v^J.
\end{equation}
Then, in matrix notation, $G = K(1 + K^{-1} M)$, so that
\begin{equation}
 \det G = \det K \det(1 + K^{-1}M)  \sim \det K( 1 + \tr (K^{-1} M) +\dots)
\end{equation}
and
\begin{equation}
 \sqrt{-G} \sim \sqrt{-K}\left( 1 + \half\tr (K^{-1} M) +\dots\right).
\end{equation}
We have dropped the $U(1)$ gauge field, which is not our concern here.
Substituting
\begin{equation}
 \sqrt{-K} = y^3 \sqrt{1+y^2} \sqrt{g}
\end{equation}
we obtain
\begin{equation}
S_1 = -\frac{N}{2\pi^2} \int  y^3 \sqrt{1+y^2}
\left(1+ \half \tr(K^{-1}M) +\ldots \right) \sqrt{g} d^3u dt.
\end{equation}
where
\begin{equation}
 \tr (K^{-1} M) = -\frac{\dot y^2}{(1+y^2)^2} - \frac{h_{IJ} \dot v^I \dot v^J}{1+y^2}
 + \frac{g^{ij} \pa_i y \pa_j y}{y^2 (1+y^2)} + \frac{g^{ij} h_{IJ} \pa_i v^I \pa_j v^J}{y^2}.
\end{equation}
The lesson we wish to emphasize is that $S_1$ contributes a potential
\begin{equation}
V_1(y) = \frac{N}{2\pi^2} y^3 \sqrt{1+y^2}.
\end{equation}

Now let us examine $S_2$. As before, $S_2 = \m_3 k_3 \int_M {\rm vol}(AdS_5)
+ \frac{\chi}{8\pi} \int F \wedge F$,
and the coefficient $\m_3 k_3$ is equal to $2N/\pi^2$. Since
\begin{equation}
{\rm vol} (AdS_5) = y^3 \wedge {\rm vol} (S^3) \wedge dt
= \frac{1}{4} d\left(y^4 {\rm vol}(S^3) \wedge dt\right),
\end{equation}
\begin{equation}
S_2 = -\int  V_2(y) \sqrt{g} d^3u dt,
\end{equation}
where
\begin{equation}
V_2(y) = -\frac{N}{2\pi^2}y^4.
\end{equation}
Altogether,
\begin{equation}
V(y) = V_1(y) +V_2(y) = \frac{N}{2\pi^2}\left(y^3 \sqrt{1+y^2} - y^4\right).
\end{equation}
Unlike the Poincar\'e-patch problem, the two terms do not cancel in this case.
In fact, $V(y)$ is a monotonically increasing function of $y$, which implies that
the D3-brane is attracted to $y =0$, the center of global AdS. It seems reasonable
that the tension of a spherical brane would cause it to collapse to a point. (In
the case of an anti-D3-brane, the sign of $V_2$ is reversed, and the collapse
force is even stronger.) The interpretation in terms of ${\cal N}=4$
nonabelian gauge theory is the following: when the theory is placed on an $S^3$,
it develops a potential that removes the Coulomb branch.
The moduli space consists of a single point.

\end{document}